\documentclass[preprints,article,accept,moreauthors,pdftex]{mdpi} 
\firstpage{1} 
\pubvolume{1}
\issuenum{1}
\articlenumber{0}
\pubyear{2022}
\copyrightyear{2022}
\datereceived{} 
\dateaccepted{} 
\datepublished{} 
\hreflink{https://doi.org/} 
\pdfoutput=1

\Title{Inhomogeneous Jets from Neutron Star Mergers: One Jet to Rule them all}

\TitleCitation{Inhomogeneous Jet Structure}

\Author{Gavin P. Lamb $^{1}$\orcidA{}*, Lorenzo Nativi$^{2}$, Stephan Rosswog $^{2,3}$, D. Alexander Kann$^{4,5}$, Andrew Levan$^{6}$, Christoffer Lundman$^{2}$ and Nial Tanvir$^{1}$}

\AuthorNames{Gavin Lamb, Lorenzo Nativi, Stephan Rosswog, D. Alexander Kann, Andrew Levan, Christoffer Lundman and Nial Tanvir}

\AuthorCitation{Lamb, G.; Nativi, L.; Rosswog, S.; Kann, D.A.; Levan, A.; Lundman, C.; Tanvir, N.}

\address{%
$^{1}$ \quad School of Physics and Astronomy, University of Leicester, University Road, Leicester LE1 7RH, UK\\
$^{2}$ \quad The Oskar Klein Centre, Department of Astronomy, Stockholm University, AlbaNova, SE-10691 Stockholm, Sweden\\
$^{3}$ \quad Hamburger Sternwarte, University of Hamburg, Gojenbergsweg 112, 21029 Hamburg, Germany\\
$^{4}$ \quad Instituto de Astrof\'isica de Andaluc\'ia (IAA-CSIC), Glorieta de la Astronom\'ia s/n, 18008 Granada, Spain\\
$^{5}$ \quad Hessian Research Cluster ELEMENTS, Giersch Science Center, Max-von-Laue-Strasse 12, Goethe University Frankfurt, Campus Riedberg, 60438 Frankfurt am Main, Germany\\
$^{6}$ \quad Department of Astrophysics, Radboud University, 6525 AJ Nĳmegen, The Netherlands}

\corres{Correspondence: gpl6@le.ac.uk}

\abstract{
Using the resultant profiles from 3D hydrodynamic simulations of relativistic jets interacting with neutron star merger wind ejecta, we show how the inhomogeneity of energy and velocity {across the jet surface profile} can alter the observed afterglow lightcurve.
We find that the peak afterglow flux depends sensitively on the observer's line-of-sight, not only via the jet inclination but also through the jet rotation:
for an observer viewing the afterglow within the GRB-bright jet core, we find a peak flux variability on the order $<0.5$ dex through rotational orientation and $<1.3$ dex for the polar inclination.
An observed afterglow's peak flux can be used to infer the jet kinetic energy, and where a top-hat jet is assumed, we find the range of inferred jet kinetic energies for our various model afterglow lightcurves (with fixed model parameters), covers $\sim 1/3$ of the observed short GRB population.
Additionally, we present an analytic jet structure function that includes physically motivated parameter uncertainties due to variability through the rotation of the source. %
An approximation for the change in collimation due to the merger ejecta mass is included and we show that by considering the observed range of merger ejecta masses from short GRB kilonova candidates, a population of merger jets with a fixed intrinsic jet energy is capable of explaining the observed broad diversity seen in short GRB afterglows.
}


\begin{document}


\section{Introduction}

The afterglows that follow short-duration Gamma-Ray Bursts (GRBs) are powered by decelerating, relativistic jets.
The likely progenitor system for short GRBs is the merger of binary neutron stars or neutron star--black hole binaries \citep[see][for a review of short GRBs]{nakar2007, berger2014, davanzo2015};
for such a compact binary merger origin, the power available for the jets is expected to be on the order of $10^{51}$ erg s$^{-1}$, with quite a narrow distribution \citep{shapiro2017, fryer2019}.
However, the jet parameters inferred via afterglow studies reveal a diverse population in terms of energy, as well as microphysics and environment \citep{fong2015, oconnor2020},
and this broad distribution in inferred kinetic energy can  be difficult to reconcile with the narrow, theoretically expected range.

A jet launched following a compact stellar merger will propagate through the merger ejecta and winds \citep[e.g.,][]{aloy2005, nagakura2014, duffell2015, murguia2017, xie2018, geng2019, nathanail2021, nativi2021, nativi2022, pavan2021, urrutia2021}, this results in the collimation of the jet before breakout \citep{bromberg2011, gottlieb2021, salafia2020, hamidani2021}.
As a consequence of the turbulent motions arising during the hydrodynamic interaction between the jet and the surrounding ejecta, the resultant jet will have an angular shape that is independent of the injected jet structure \citep{nativi2022}, unless the ejecta density is very low or the jet power very high \citep{urrutia2021}.
\footnote{The different initial conditions between these two studies make a direct comparison non-trivial; whereas \citet{nativi2022} use initially ``warm'' jets, \citet{urrutia2021} have initially ``cold'' jets and include a broader jet luminosity distribution.
Additionally, we note that the turbulent effects seen in 3D simulations are significantly reduced in 2D \citep[see][]{matsumoto2019a}; thus the turbulent destruction of the initial jet structure seen in the 3D hydrodynamic simulations by \citet{nativi2022} is not as pronounced in the 2D hydrodynamics used by \citet{urrutia2021}.}
However, unlike the functional form of the resultant jet structure profile due to this turbulence, the characteristic opening angle of the emergent jet can retain information about the injected jet's angular width \citep[e.g.,][]{nagakura2014}.

A universal, or quasi-universal, jet structure profile for GRB jets with a standard energy reservoir was first proposed by \citet{lipunov2001} and expanded upon by \citet{rossi2002, zhang2002}.
For such universal structured jet models, the broad range in observed GRB energies is a result of the arbitrary viewing angle.
Analysis of the spectral peak energy to isotropic energy correlations for a population of such universal GRBs can be used to constrain the viable jet structure profile \citep[e.g.][]{lloyd-ronning2004, dai2005, lamb2005}.
For long GRBs, classically associated with a collapsar origin, the luminosity function can be used to exclude some shallower jet structure profiles \citep{pescalli2015}, however, empirical correlations may well be a result of an apparent jet structure \citep{salafia2015}, and consistent with the resultant structure profiles from simulations of jets in massive stars \cite[e.g.,][]{xie2019}.
Additionally, observational signatures for the presence of jet structure within GRB-producing jets are contained within the various afterglow lightcurve decline rates and breaks \citep{kumar2003, rossi2004, takami2007, lamb2021}.

As GRBs are highly beamed, their afterglows are preferentially selected to be at small inclination angles to the line-of-sight, $\iota$, where the emission is brightest \citep{beniamini2019}.
For such GRBs, the afterglow lightcurve can be modelled using a simple top-hat jet structure, where any angular dependence of the energy or velocity is ignored, resulting in a uniform jet within a cone defined by the jet's opening angle, $\theta_c$ \citep[see discussion in][]{aksulu2021}.
However, where a top-hat jet structure model is assumed, the inferred energy from lightcurve fits will return kinetic energy values equal to the average energy of the intrinsic jet structure profile within the $1/\Gamma$ beaming cone and likely not the true or maximum energy of the jet.
Typically, $1/\Gamma < \theta_c$ at peak flux time where the observer's line-of-sight is $\iota \lesssim \theta_c$ -- here $\Gamma$ is the bulk Lorentz factor of the emitting region at the observation time.
This should not be confused with higher inclination cases, where the GRB is either absent or at a much lower luminosity e.g., the expected afterglow counterparts to gravitational wave detected mergers \citep[see][for details of such ``off-axis'' observed afterglows]{lamb2017, lazzati2017}.
Here we use the results of 3D hydrodynamic simulations of jets propagating through the merger ejecta and neutrino-driven winds of a neutron star merger \citep[see,][]{perego14,nativi2021, nativi2022} to investigate the effects of inhomogeneity within the jet's energy and velocity profile on the afterglows of short GRBs.

In \S\ref{sec:method} we describe our method for modelling the resultant jet profiles from the simulations relevant to the bright GRB emitting population of short bursts.
In \S\ref{sec:results} we show the diversity of afterglow lightcurves from a single jet simulation as a function of observer line-of-sight relative to both inclination and rotation.
Additionally, we use the simulation jet structure results to generate a general, analytic function for the typical  structure of a short GRB jet and  compare the flux density for the afterglow from this model to the short GRB afterglow population.
These results are discussed in \S\ref{sec:disc}, where we approximate how the ejecta mass responsible for a thermal kilonova \citep[e.g.,][]{rosswog1999,hotokezaka13} can alter the effective core-size of the jet and show how this compares to a selection of typical afterglow lightcurves for short GRBs with candidate kilonovae.
Our conclusions are listed in \S\ref{sec:conc}.

\section{Method}\label{sec:method}

The energy per steradian and Lorentz factor at each point of a surface that describes the afterglow-producing jet can be extracted from simulations.
We use two ultra-relativistic jet simulations from \cite{nativi2022}, and determine the typical Lorentz factor for each surface element from the mass-averaged radial profile for $h\Gamma>2$ at a given polar angle, $\theta$, and rotational angle, $\phi$ -- where $h$ is the specific enthalpy, and $\Gamma$ the Lorentz factor.
The two simulations we utilise are identical but for the structure of the injected jet in each case; the first uses a `top-hat' structured jet with a uniform energy and enthalpy until a sharply cut-off edge, and the second uses a profile described by a Gaussian function. 
Both jets have a power, $L_{\rm j} \sim 10^{50}$ erg s$^{-1}$, and are labelled \texttt{th50} and \texttt{gs50} respectively.

\begin{figure}[H]
    \begin{adjustwidth}{-\extralength}{0cm}
    \centering
    \includegraphics[width=18cm]{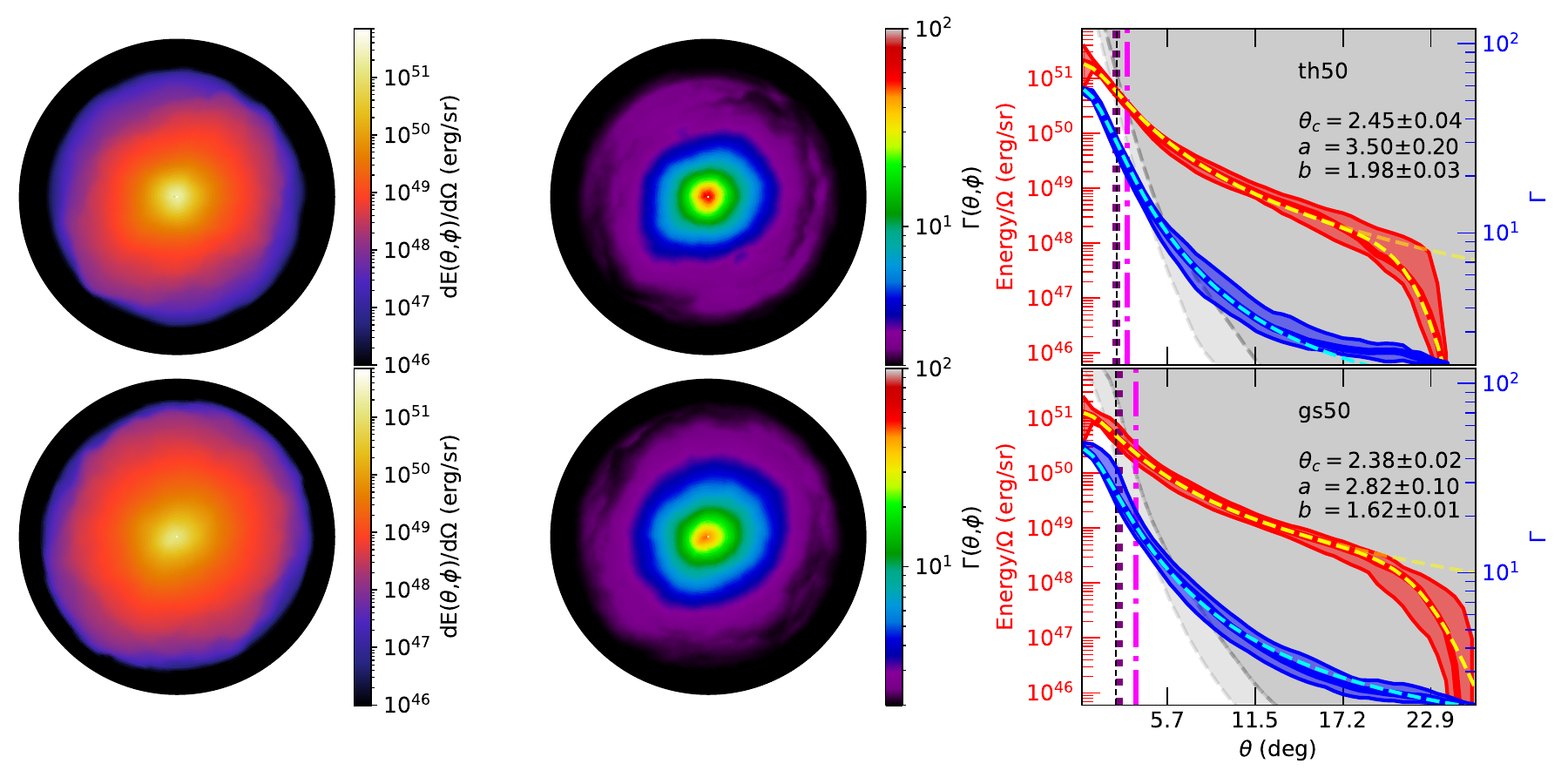}
    \end{adjustwidth}
    \caption{[{ Polar-plots}] Face-on projections of the resultant jets/outflows. From the centre, $\theta=0$, to the edge, $\theta=29^\circ$.
    Each row indicates a different initial jet structure described by: \texttt{th50, gs50}, top to bottom.
    Where \texttt{th} indicates a top-hat profile for the injected jet structure, \texttt{gs} indicates a Gaussian structure, and the number is the logarithm of the jet power i.e., $10^{50}$ erg s$^{-1}$.
    [{ Left}]: The energy per steradian (erg sr$^{-1}$) for the resultant jet profiles. 
    Energy per steradian is shown in the range $1\times10^{46}\leq {\rm E}/\Omega \leq 7\times10^{51}$ erg sr$^{-1}$.
    [{ Centre}]: The resultant mass-averaged Lorentz factor for the same projection.
    Bulk Lorentz factor is shown in the range $2\leq\Gamma\leq100$.
    [{ Right}] The maximum and minimum energy per steradian (red) and Lorentz factor (blue) in rotation at each polar angle $\theta$ from the central axis for the model. 
    The largest logarithmic variations are seen at wider angles where energies and Lorenz factors are lower -- the average value is shown as a thick line within the shaded region.
    The purple, vertical dotted line indicates the jet opening angle as inferred by the afterglow break time for an on-axis observer -- the black dashed line is the core opening angle found via fits to the mean angular profile, showing good agreement with the values inferred via the jet break \citep{nativi2022}.
    The grey shaded area indicates the region within which we don't expect detectable $\gamma$-ray emission due to opacity arguments, see equation \ref{eq:energy}, where we have assumed $\eta=0.15$, $T_{\rm dur} = 0.1$s, and $\delta t = 0.1$s -- the lighter grey region indicates the limit using the minimum $\Gamma$ value, while the darker region indicates the limit using the maximum $\Gamma$ value for each angular segment.
    The pink, dash-dotted line indicates the maximum angle for $\gamma$-ray emission considering only the opacity due to electrons that accompany baryons in the jet.
    The yellow dashed line indicates the approximate functional shape of the jet profile in terms of energy with the fainter line at wider angles showing the profile without the energy cut-off, while the cyan dashed line represents the Lorentz factor profiles -- the fit values for $\theta_c$, $a$ and $b$ in each panel are those for the analytic function in equations \ref{eq:sj1}--\ref{eq:sj2} (see text for details).}
    \label{fig:polar}
\end{figure}

The two left panels in Figure \ref{fig:polar} show the face-on distribution of energy and Lorentz factor for each simulation output once ballistic expansion is achieved i.e., $(h-1)<1$ everywhere on the grid.
The right panels show energy (red) and Lorentz factor (blue) with polar angle where the shaded regions indicate the maximum to minimum range for each parameter with rotation through $\phi$ on the jet surface at the given angle $\theta$.
We can approximate the mean of these profiles analytically using \citep[e.g.][]{beniamini2020}:
\begin{eqnarray}
    \Theta &=& \left[1+\left(\frac{\theta}{\theta_c}\right)^2\right]^{1/2},\label{eq:sj1}\\
    E(\theta) &=& E(\theta=0)\Theta^{-a},\label{eq:sj3}
    \\
    \Gamma(\theta) &=& 1+\left[\Gamma(\theta=0)-1\right]\Theta^{-b}.\label{eq:sj2}
\end{eqnarray}
We use a linear regression fit to the mean for both the energy and Lorentz factor profiles within an angle $\theta_j = 0.42$ rad, or $\sim24^\circ$.
The energy profile is cut-off\footnote{The wide angle cut-off in energy for the simulation profiles is a result of the radial averaging process where we only sample components with $h\Gamma > 2$. The contribution to the observable afterglow from wider, lower energy and $\Gamma$ regions is negligible, see \citet{nativi2022} for details.} using the functional form $(1+(\theta/\theta_j)^{a_2})^{-a_1}$.
The model jet structure (equations \ref{eq:sj1}--\ref{eq:sj2}) fit parameters to the mean through rotation for the simulation profiles are shown in Table \ref{tab:profiles}.

\begin{table}[H]
    \caption{Analytic jet structure profile parameters, see equations \ref{eq:sj1}--\ref{eq:sj2} plus text, from fits to the mean energy and Lorentz factor with polar angle for each simulation and the averaged profile, see \S\ref{sec:results}.}
    \label{tab:profiles}
    \begin{adjustwidth}{-\extralength}{0cm}
	\newcolumntype{C}{>{\centering\arraybackslash}X}
	\begin{tabularx}{\fulllength}{c|c|c|c|c|c|c|c}
        Model & $\theta_c$ (rad) & $\log(E_c)$ ($\log$ erg sr$^{-1}$) & $\Gamma_c$ & $a$ & $a_1$  & $a_2$ & $b$ \\
        \hline
        \texttt{th50} & $0.0428\pm0.0007$ & $51.27\pm0.11$ & $58.0\pm0.2$ & $3.50\pm0.20$ & $7.89\pm0.94$ & $14.79\pm3.14$ & $1.98\pm0.03$\\
        \texttt{gs50} & $0.0415\pm0.0004$ & $51.10\pm0.06$ & $45.6\pm0.1$ & $2.82\pm0.10$ & $3.96\pm0.45$ & $12.48\pm2.59$ & $1.62\pm0.01$\\
        Averaged & $0.0424\pm0.0005$ & $51.18\pm0.08$ & $52.3\pm0.2$ & $3.10\pm0.14$ & $3.64\pm0.54$ & $10.90\pm3.02$ & $1.82\pm0.02$
        \end{tabularx}
    \end{adjustwidth}
\end{table}

To test how the inhomogeneity of the jet surface affects the observed afterglow lightcurves for various lines of sight, we generate afterglows at a fixed emission frequency for observers at different combinations of polar and rotational angles.
The afterglows are calculated using the method described in \cite{lamb2017, lamb2018, lamb2021}.
We use the energy and Lorentz factor for each surface element from the simulation output and calculate the afterglow contribution from each, as seen by an observer at a given inclination, $\iota$, and observer time, $t_{\rm obs}$, and sum the equal time contributions to give the total observed flux with observer time i.e., we integrate over the equal arrival time surface.
This is the current standard for structured jet afterglow models \cite[e.g.,][]{lamb2018, ryan2020}, however, we note that this assumes the jet surface profile is frozen until lateral expansion begins.
As the outflow has a very low enthalpy i.e., $(h-1)<1$, at the time we sample the jet profile, the frozen-in assumption is a reasonable approximation until the outflow begins deceleration \citep[see also][]{wang2021}. 
Hydrodynamic simulations of relativistic jets show that lateral spreading only becomes significant at late times during the afterglow emission phase, \cite[e.g.][]{vaneerten2013}, thus we expect no significant changes to the jet surface profile until very late times

\begin{figure}
    \centering
    \includegraphics[width=\columnwidth]{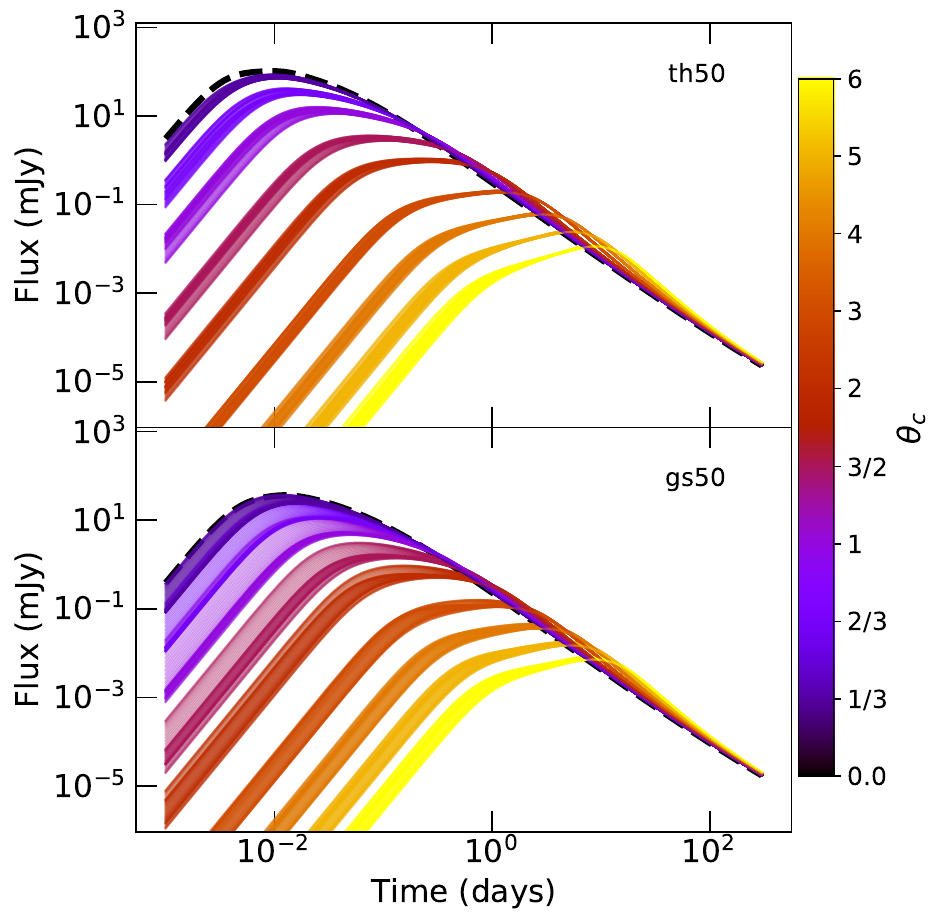}
    \caption{Lightcurves {in $i$-band} at various polar angles, $\theta$, from the jet {central} axis. Here the polar angle is equivalent to the system inclination,  $\theta \equiv \iota$, and shown in the colour bar as a fraction of the effective core angle for each profile, where the line colour indicates the relative angle in terms of core size -- the discrete angles shown correspond to the colour bar labels. The effective core angles {for each model, $\theta_c$,} are given in Table \ref{tab:profiles}. At each $\theta \equiv \iota$, lightcurves at all available rotational orientations, $0\leq\phi\leq2{\pi}$, are shown in the same colour. {Here} the apparent broadness of the lines {is indicative of} the spread in flux as a result of the rotational orientation{, $\phi$, of the jet to the line-of-sight}.
    For a fixed inclination angle, the rotational orientation of the system has a significant effect on the flux density at $\sim$peak time, or equivalently the deceleration time, for an observer $\iota\lesssim\theta_c$. Where $\iota\gtrsim2\theta_c$, then only the early and pre-peak flux density is sensitive to the rotational orientation.
    The dashed line indicates the $\iota = 0$, perfectly on-axis afterglow lightcurve.}
    \label{fig:prelim-lc}
\end{figure}

The lightcurves for each $\phi$ element at a discrete polar angle in the range $(0.0 \leq \theta \leq 6.0) \times\theta_c$ for the two simulation models are shown in Figure \ref{fig:prelim-lc}.
The lightcurves are calculated at an observed frequency, $\nu=3.8\times10^{14}$\,Hz, where for simplicity we assume redshift, $z=0$, and a luminosity distance, $D_L=100$\,Mpc.
The afterglow model microphysical parameters are fixed at $\varepsilon_e = \sqrt{\varepsilon_B}=0.1$, an electron distribution index, $p=2.15$, an ambient medium particle number density, $n=1$\,cm$^{-3}$, and a $\gamma$-ray efficiency of $\eta=0.15$, where we assume a fixed efficiency throughout the jet.

The emission of $\gamma$-rays from a jet requires the source to be optically thin which places physical constraints on the angular profile from which a GRB can be emitted/observed \citep[e.g.][]{lamb2017}. 
Viable locations for $\gamma$-ray emission can be determined using a relation between energy and Lorentz factor that considers the opacity of the medium to gamma-rays.
The minimum Lorentz factor for an optically thin medium considering only the scattering by electrons that accompany baryons in the jet is \citep[e.g.][]{lithwick2001, matsumoto2019}:
\begin{equation}
    \Gamma_{\rm min} = \left(\frac{\sigma_{\rm T}}{32 m_{\rm p}{\pi} c^4}\frac{L_{\gamma,{\rm iso}}(1+z)}{\delta t}\right)^{1/6},
    \label{eq:opacity}
\end{equation}
which is consistent with $\Gamma_{\rm min}\propto {E_{\gamma,{\rm iso}}}^{0.17}$ found in \cite{lamb2016}.
Here, $\sigma_{\rm T}$ is the Thompson cross-section, $m_{\rm p}$ is the mass of a proton, $c$ is the speed of light, and the variables: $L_{\gamma,{\rm iso}}$ is the isotropic equivalent $\gamma$-ray luminosity, $z$ is the source redshift, and $\delta t$ is the minimum variability timescale.
The observed isotropic equivalent $\gamma$-ray energy can be approximated from the luminosity as, $E_{\gamma,{\rm iso}} = L_{\gamma,{\rm iso}} T_{\rm dur}$, where $T_{\rm dur}$ is the burst duration\footnote{For an aligned observer, $\iota=0$, the GRB duration is equivalent to the engine duration; for our simulation this is $0.1$s}.
The total energy is then $E = E_{\gamma,{\rm iso}}/\eta$, where $\eta$ is the efficiency of the $\gamma$-ray emission.
A basic approximation of the 
energy per steradian for the outflow with a given $\Gamma$ to be $\gamma$-ray bright is then
\begin{equation}
    \frac{E}{\Omega} \lesssim  \left(\frac{\Gamma}{10}\right)^6 \left(\frac{\delta t}{0.1 {\rm s}}\right) \left(\frac{T_{\rm dur}}{0.1 {\rm s}}\right)  \left(\frac{\eta}{0.15} \right)^{-1}\times 10^{48}~ {\rm erg/sr},
    \label{eq:energy}
\end{equation}
where all timescales are measured in the lab frame.
{Equation \ref{eq:energy} provides a conservative estimate for the jet regions that can emit GRBs and the dependence on the burst duration e.g., engine timescale, minimum variability timescale, and $\gamma$-ray efficiency.}

\section{Results}\label{sec:results}

\begin{figure}
    \centering
    \includegraphics[width=\columnwidth]{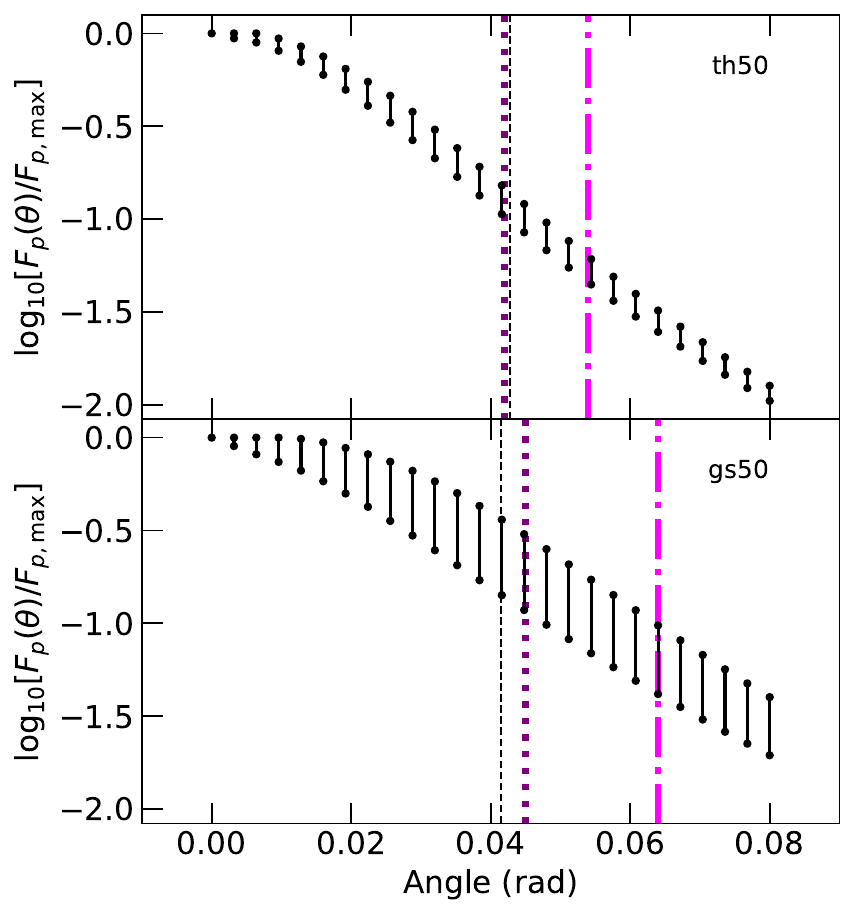}
    \caption{The range in afterglow peak flux values relative to the maximum afterglow peak flux through rotation in $\phi$ of the jet at a fixed polar angle{ , equivalent to the line-of-sight inclination,} within the $\gamma$-ray bright jet region for each structure profile.
    The purple dotted vertical line indicates the inferred jet opening angle for an on-axis observer, given the jet break time; the black dashed line indicates the jet core angle from mean jet profile fits; and the pink dash-dotted line indicates the maximum angle at which the jet is optically thin to $\gamma$-rays.}
    \label{fig:DLFp}
\end{figure}

We have generated afterglow lightcurves from the resultant energy and Lorentz factor surface profiles for two 3D hydrodynamic simulations of jets propagating through neutron star merger winds for observers at various $\theta$ and $\phi$ relative to the jet central axis \citep[see][for simulation details]{nativi2021, nativi2022}.
The effects of the jet orientation to the line-of-sight on the peak afterglow flux, for emission in the regime $\nu_m<\nu<\nu_c$, where $\nu_m$ is the characteristic synchrotron peak frequency and $\nu_c$ is the cooling frequency, is shown in Figure \ref{fig:DLFp} for observers within the $\gamma$-ray emitting region of the jet (as defined by equation \ref{eq:energy}).
The maximum variation in the peak flux at a fixed inclination but through a $2\pi$ rotation in $\phi$ is $\sim0.5$ dex seen for the \texttt{gs50} model.
The \texttt{th50} model has less overall variation, with $\sim0.2$ dex in peak afterglow flux.
The most significant change in peak flux is seen with inclination from the jet central axis, where for both \texttt{th50} and \texttt{gs50} the peak flux varies by $\sim1.3$ dex within the approximate $\gamma$-ray emitting region.

\begin{figure}
    \centering
    \includegraphics[width=\columnwidth]{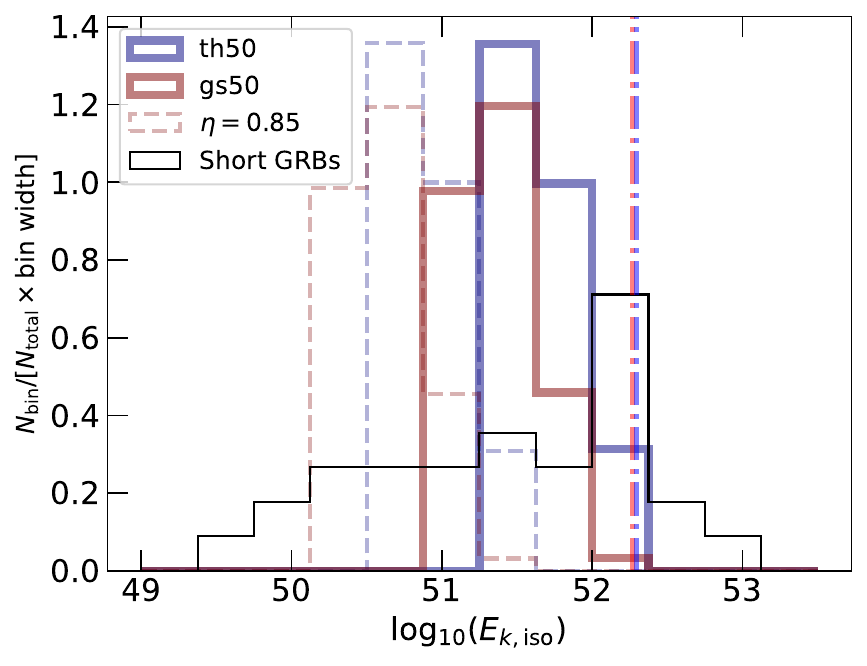}
    \caption{Probability density of the kinetic energy from our model jet structure profiles inferred from the peak afterglow flux, versus the observed short GRB kinetic energy distribution as listed in \citet{fong2015} for $\varepsilon_B=0.01$. For a fixed $\gamma$-ray efficiency ($\eta=0.15$), a single jet profile can account for $\sim$one third of the observed logarithmic kinetic energy distribution range -- the distribution with a fixed $\eta=0.85$ is shown as a dashed line.
    The dash-dotted lines indicate the isotropic equivalent energy for each jet assuming the intrinsic jet energy, $E_j\sim10^{49}$ erg, is contained within a cone defined by the jet structure profile core angle.}
    \label{fig:ekinetic}
\end{figure}

For GRB afterglows, the kinetic energy of the jet is found via afterglow modelling which typically assumes an outflow with a uniform, or top-hat, energy distribution \citep[e.g.][]{fong2015}.
The results of our simulations show that the jets that produce GRB afterglows do not have a uniform energy profile, and that orientation through both $\theta$ and $\phi$ can change the kinetic energy inferred from simple afterglow models.
For GRB afterglows with $\nu_m<\nu<\nu_c$ the peak flux is $F_p\propto {E_{\rm k}}^{(3+p)/4}$, where $F_p$ is the afterglow peak flux, $E_{\rm k}$ is the kinetic energy, and $p$ is the power-law distribution index for accelerated electrons \citep{sari1998} -- note that we ignore the dependence on the ambient medium density, $n$, and the microphysical parameters, $\varepsilon_B$ and $\varepsilon_e$ as these quantities are fixed for our sample.
Figure \ref{fig:ekinetic} shows the kinetic energy distribution, as inferred by a distant observer and weighted for a randomly oriented source with inclination $\iota\leq\theta_\gamma$, for each of our jet simulations in comparison to the observed kinetic energy distribution for the population of short GRBs as listed in \cite{fong2015}.
The afterglow lightcurves were calculated assuming a fiducial efficiency for $\gamma$-rays of $\eta=0.15$, where the energy in the jet is $E = E_\gamma + E_{\rm k}$, with $E_\gamma$ being the energy radiated in $\gamma$-rays.
The dashed histogram shows the distribution assuming an efficiency, $\eta=0.85$, resulting in a lower typical energy distribution as more energy is lost via the GRB.
The dash-dotted lines show the isotropic equivalent kinetic energy for each of our jets, calculated assuming the core angle listed in Table \ref{tab:profiles} contains all of the initial jet energy. 
The logarithmic kinetic energy distribution inferred from a single jet model covers $\sim$third of the observed short GRB population.
As we have artificially fixed the $\gamma$-ray efficiency for the entire emitting region, our distribution is likely significantly narrower than one with a more realistically determined efficiency e.g., $\eta(\theta,\phi)$, that varies according to local conditions across the jet's surface.

\begin{figure}
    \centering
    \includegraphics[width=\columnwidth]{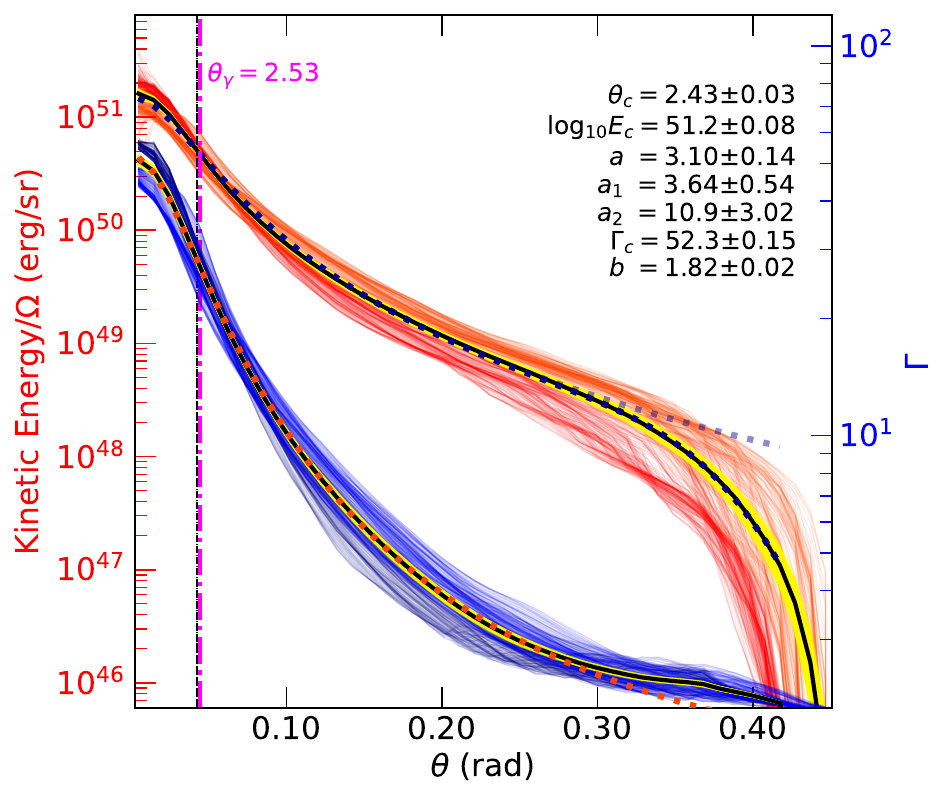}
    \caption{The energy (left axis) and $\Gamma$ (right axis) profiles at each rotation step in $\phi$ for both the \texttt{th50} (red/dark-blue) and \texttt{gs50} (orange/blue).
    By assuming that the differences in these model profiles are the result of chaotic mixing processes while the jet propagates through the neutron star merger ejecta, we use a bootstrap to find the average profile (yellow are individual bootstrap means, black gives the distribution mean).
    We then fit our analytic jet structure profile for $0.0\leq\theta\leq0.42$\,rad to the distribution mean -- the analytic model fit parameters are shown in the figure, and the profiles are indicated with a dotted line for both the energy and $\Gamma$.}
    \label{fig:bootstrap}
\end{figure}

The two 3D simulations highlight that the form of the resultant jet profiles are largely a result of fluid instabilities in the jet-wind interaction regions  \citep{nativi2022}, however, for lower density winds\footnote{The density and mass of the merger winds in the simulations of \cite{nativi2021, nativi2022} are already low and the emergent jet structure did not preserve the injected profile. We do not expect many physical scenarios where the injected jet's structure contributes significantly to the emergent jet profile.}, or much more powerful jets, the injected jet structure can be partially preserved \citep{urrutia2021}.
By using the rotation in $\phi$ for each surface to produce an energy and Lorentz factor profile in $\theta$ we can find an average jet structure profile from our simulations that includes the rotational variation seen here.
{We use a bootstrapping technique, re-sampling these unique jet profiles, to produce a sample of mean profiles and then find the mean of this new sample.}
Equations \ref{eq:sj1}--\ref{eq:sj2} {are fit to this mean profile}, {including} the energy cut-off described in \S\ref{sec:method}, via linear regression to find the best-fit parameters.
These are listed in Table \ref{tab:profiles} as `Averaged' and shown in Figure \ref{fig:bootstrap} along with the initial profiles.

\begin{figure}
    \centering
    \includegraphics[width=\columnwidth]{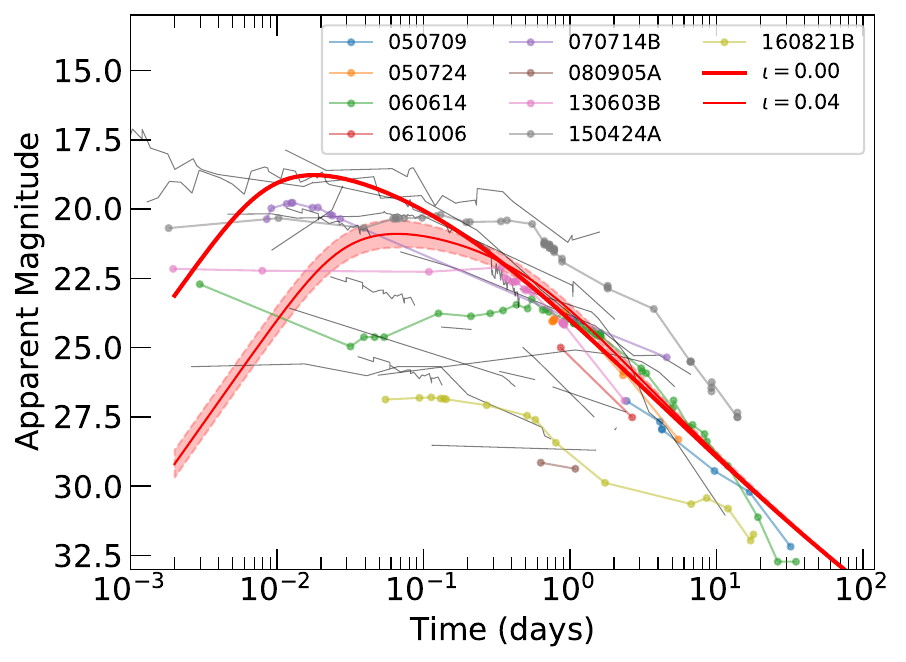}
    \caption{The averaged model afterglow lightcurve range for observers within the $\gamma$-ray bright angle, $\theta_\gamma$; red lines indicate the afterglow viewed at an inclination $\iota =$ 0, and the widest angle for $\gamma$-ray emission, where the shaded region includes the maximum variability in peak flux expected due to the rotational orientation of the jet. 
    The short GRB afterglows within our sample are shown as thin grey lines, and coloured lines with points marking data for short GRBs with candidate kilonovae.
    All lightcurves have been normalised to a redshift, $z=1$, for consistency.
    Whereas the model lightcurves have a fixed ambient medium, $n = 1$\,cm$^{-3}$, much of the diversity in the observed short GRB afterglows can be accounted for by varying ambient medium density values between events.
    This suggests that the intrinsic energy distribution of short GRB jets is quite narrow with environmental effects determining much of the population scatter.}
    \label{fig:Kannplot}
\end{figure}

The afterglows from our averaged jet structure profile are compared to the observed short GRB population in Figure \ref{fig:Kannplot}.
Here we show, in red, the model $R_C$-band afterglow for a source at $z=1$ using the same fiducial parameters as the earlier models.
Individual lightcurves are shown for an observer that is either aligned with the jet central axis, or at $\iota=\theta_\gamma$, the maximum angle at which $\gamma$-rays are emitted for our model; these cover the range of expected GRB afterglows from a jet with our structure and fixed parameters.
The sample of 30 individual short GRBs (grey and coloured lines) are optical afterglows for bursts with a measured redshift;
each afterglow is a composite of various filters that have been shifted using the spectral energy distribution for each burst and corrected for Galactic foreground extinction and host contribution (if necessary and possible) to produce an observed $R_C$-band lightcurve for a source at redshift $z=1$ \citep[see][and references therein]{kann2011, agui2021}.

The post-jet-break decline for our model lightcurves is consistent with the tail of the short GRB population, and the peak of the model afterglows agrees nicely with the brightest in this distribution -- we note that short GRB afterglows shown here may include contributions from extended prompt emission, the reverse shock, energy injection, and kilonova.
Our afterglow models have a fixed ambient density, $n = 1$ cm$^{-3}$, and microphysical parameters, $\varepsilon_e = \sqrt{\varepsilon_B}=0.1$ and $p=2.15$; allowing these to vary would change both the timescale and the peak flux of the afterglow, with the deceleration (or peak) time, $t_d \propto {E_{\rm k}}^{1/3} n^{-1/3} \Gamma^{-8/3}$, and $F_p \propto E_{\rm k} \Gamma^{2(p-1)} {\varepsilon_B}^{(1+p)/4} {\varepsilon_e}^{p-1} n^{(1+p)/4}$, where $\Gamma$ is the bulk Lorentz factor at the deceleration radius and the emitting frequency is in the regime, $\nu_m<\nu<\nu_c$.

\section{Discussion}\label{sec:disc}

We have used the results from 3D hydrodynamic simulations of relativistic jets interacting with neutron star merger winds to show the effect on the observed afterglow, in terms of the peak flux, from the inhomogeneity of the resultant jet structure in both polar, $\theta$, and rotational, $\phi$, orientations.
We have shown that for an observer viewing the jet at $\iota\leq\theta_\gamma$, the rotational orientation of the jet surface, $0\leq \phi \leq 2{\pi}$ results in an afterglow peak flux with a scatter of $\Delta\lesssim 0.5$ dex.
The structure through the polar-angle, $0\leq\iota\leq\theta_\gamma$, results in a larger scatter for the peak afterglow flux of $\lesssim 1.3$ dex, as expected from previous studies of jet structure on the afterglows to merger jets which only considered the polar variation in jet properties \citep{lazzati2017, lamb2017}.
For the scatter in flux density due to the rotational orientation, the deviation from the mean gradually reduces with time and follows the expected change in beaming angle; the flux changes by a factor $\lesssim 1+(\Delta/2-1)\times \min[{1,(t_d/t)}]^{3/8}$, where  $t_d$ is the deceleration time, and $t$ is the observer time since burst.

The jet's kinetic energy, as inferred via afterglow modelling, will be broader than that assumed from a simple uniform jet model.
For our two fiducial models, we find the inferred kinetic energy distribution covers $\gtrsim 1$ dex in energy for a fixed $\gamma$-ray efficiency, $\eta$.
More realistically, $\eta$ would vary as a function of both the energy and Lorentz factor with $\theta$ and $\phi$, where at lower Lorentz factors some fraction of the energy dissipated within the jet will be reabsorbed \citep[see,][]{kobayashi2001, kobayashi2002} and the effective $\eta$ would be smaller.
Higher energy regions may result in more efficient shocks \citep[e.g.][]{kobayashi2001} and the $\eta$ could be larger -- similarly, for photospheric emission during the prompt emission of GRBs, \citet{, gottlieb2019} find a higher efficiency in more energetic regions.
Such properties would further broaden the inferred kinetic energy distribution from a jet with fixed energy and we encourage further investigations into a physically motivated expression for $\eta(E,\Gamma)$ in GRB outflows.

The core angle, $\theta_c$, from the fits to the mean energy and Lorentz factor profiles for equations \ref{eq:sj1}--\ref{eq:sj2}, see Table \ref{tab:profiles}, returns a remarkably similar value to that inferred via modelling the afterglow jet-break time in \cite{nativi2022} using the same simulation data but a rotationally averaged jet profile.
For jet simulations such as these, the core or jet opening angle is often presented as the average angle within which material has $h\Gamma>10$ \citep[e.g.][]{nagakura2014}.
Using this method with our 3D simulations, we find a $\theta_{\rm average} = 0.1058$ and $0.1278$ rad for the \texttt{th50} and \texttt{gs50} models respectively.
These are both larger than the values found via profile fitting or the jet-break time by a factor $\sim 2.5$--$3$.
This suggests that the apparently narrower jets from our simulations are merely a result of the way $\theta_c$ is estimated.
Such narrow jets are well within the observed range for opening angles\footnote{The opening angle inferred via the jet-break time of a GRB afterglow typically assumes that the observer is on the jet central axis. More detailed studies of the GRB population indicate that the typical inclination for an observed GRB is $0.57$ of the jet's effective opening angle, $\theta_c$ in our notation, \citep{ryan2015}. This suggests that the jet opening angle for GRBs are typically smaller by a factor $\sim0.64$ than the simple estimates \citep{lamb2021}.} inferred from the short GRB population \citep[e.g.,][]{jin2018, lamb2019}.

The degree of the jet collimation for a GRB producing jet, and the resultant jet opening angle is a complicated function of the jet's power, the density of the medium through which the jet is propagating, and the jet's initial opening angle \citep{bromberg2011}.
The details of how to estimate the final jet opening angle from a set of initial conditions and for an expanding medium are described by \cite{hamidani2021}, \citep[see also][etc.]{murguia2017, salafia2020}.
The degree of collimation is shown to be dominantly proportional to the ratio of jet power and ambient medium mass, $\theta/\theta_0 \propto (L_j/M_{\rm a})^{1/4}$, where $\theta$ is the resultant opening angle, $\theta_0$ is the injected jet's opening angle, $L_j$ is the jet power, and $M_{\rm a}\equiv M_{\rm ej}$ is the mass of the ambient medium/winds through which the jet is propagating and equivalent to the ejecta mass for a neutron star merger.
The exact relation depends on how the medium is expanding, the ratio of the energy density for the jet to ejecta, and weakly on the timescale.
Additionally, particle effects such as neutron conversion-diffusion may contribute to the resultant jet structure \citep{preau2021}.

\begin{figure}
    \centering
    \includegraphics[width=\columnwidth]{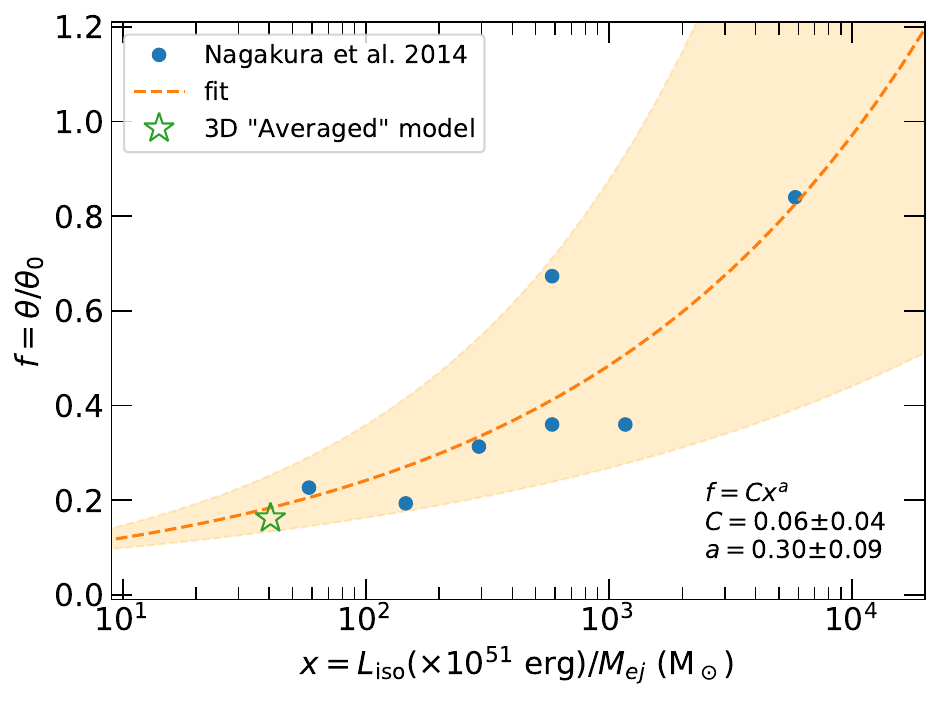}
    \caption{A fit to the ratio of jet power to ejecta mass and the degree of collimation for GRB jets from neutron star mergers.
    We use the results from 2D simulations in \citet{nagakura2014} (blue circles) for our fit (orange dashed line), and show where our averaged fiducial jet profile fits on this plot (green star).
    The shaded region shows the $a=0.30\pm 0.09$ range.
    Note -- our fiducial jet was not used in the fitting process.}
    \label{fig:collimation-rel}
\end{figure}

The afterglows to short GRBs indicate that they are accompanied by a broad diversity of kilonovae \citep[e.g.][]{gompertz2018, rossi2020}. 
Model fits to GRB afterglows with kilonova candidates suggest that the population ejecta mass distribution has a scatter of $\Delta \log M_{\rm ej}\lesssim 1$, a broad uncertainty on the ejecta velocity, $v_{\rm ej}\leq0.3$c, and a several orders of magnitude range for the Lanthanide fraction \citep{ascenzi2019}.
Theoretical studies of kilonova models show an equally diverse range of possible parameters \citep[e.g.][]{kawaguchi2020}.
The mass and velocity of the ejecta, and the power and initial opening angle of the jet have an influence on the resulting jet opening angle -- to approximate these factors, we take the results of 2D hydrodynamic simulations for a variety of initial jet and ejecta compositions from \cite{nagakura2014} and fit a power-law function, $\theta/\theta_0 \propto(L_j/M_{\rm ej})^a$, to determine the index $a$ for the scaling.
The results are shown in Figure \ref{fig:collimation-rel}, where we have used the isotropic equivalent luminosity for the initial jet.
The opening angle, as inferred from jet profile fitting, for our averaged jet structure model is shown as a star.
The fit index $a=0.30\pm0.09$ is within error of the theoretically expected, $a=0.25$, where we ignore the timescale and ejecta expansion dependence \citep{hamidani2020}.  

Using the fit {for the scaling relation $\theta_c/\theta_0\propto(L_j/M_{\rm ej})^a$} shown in Figure \ref{fig:collimation-rel}, the core angle for a structured jet with our `Averaged' jet profile, as a function of the jet energy and{/or} the ejecta mass, is
\begin{equation}
    \frac{\theta_c}{\theta_0} = 0.1696\left[\frac{E_{j}/1.7\times10^{49}~{\rm erg}}{M_{\rm ej}/0.072 ~M_\odot}\right]^{0.3}~~~{\rm rad},
    \label{eq:theta_fit}
\end{equation}
where $E_{j}$ is the equivalent energy for the jets { calculated assuming the isotropic equivalent core energy at $\theta=0$ and a top-hat jet with an opening angle $\theta_j \equiv \theta_c$ e.g., $E_j = E_{\rm iso}(\theta=0) \theta_c^2/2$, thus ensuring that the jet core energy is uniform between cases,} $M_{\rm ej}$ is the mass of the ejecta/wind, and $\theta_0$ the initial (injected) jet's half opening angle (in our simulations $\theta_0 = 0.25$\,rad).

\subsection{Comparison to sources with kilonova candidates}
Figure \ref{fig:Kannplot} shows our sample of 30 composite short GRB afterglows normalised to $z=1$ \citep{kann2011,agui2021}. The GRBs with kilonova candidates within this sample have coloured lines, points marking the data, and a corresponding legend identifier.
These kilonova candidate GRBs are additionally listed in Table \ref{tab:KN-GRBs}, with the literature values for merger ejecta masses inferred from model fits/estimates to observations by the listed reference.

\begin{table}
    \caption{Short GRBs with kilonova candidates. Estimated ejecta masses are from the respective reference column -- where more than one mass is listed, the values are in the reference order; the bold value is that used to determine the jet collimation for the models shown in Figure \ref{fig:LCs_AG}. GRBs 150101B and 200522A, below the line, are not included in our optical lightcurve sample. The 4th and 5th columns show the value of index $p$ used for the model afterglow lightcurves, and the relevant reference. The last three columns show the opening angle, the isotropic equivalent kinetic energy, inferred from the jet structure in each case, and the ambient particle density used for the lightcurves in Figure \ref{fig:LCs_AG}. The ejecta mass ($M_{\rm ej}$) is in units M$_\odot$, core angle ($\theta_c$) is in radians, kinetic energy ($E_{\rm k}$) is in erg, and the ambient density ($n$) in  cm$^{-3}$.}
    \label{tab:KN-GRBs}
	\newcolumntype{C}{>{\centering\arraybackslash}X}
	\begin{tabularx}{\textwidth}{c|c|c|c|c|c|c|c}
            GRB & $M_{\rm ej}$ & Ref. & $p$ & Ref. & $\theta_c$ & $E_{\rm k}$ $\times10^{52}$ & $\log n$\\
            \hline
            050709 & {\bf 0.05} & \cite{jin2016} \cite{ascenzi2019} & 2.31 & \cite{fong2015} & 0.047 & 1.29 & -3.00 \\
            050724 & {\bf 0.001} & \cite{gao2017} & 2.29 & \cite{fong2015} & 0.160 & 0.11 & -1.00 \\
            060614 & {\bf 0.077}, 0.1 & \cite{ascenzi2019} \cite{jin2016} & 2.40 & \cite{jin2015} & 0.042 & 1.69 & -3.30 \\
            061006 & {\bf 0.01} & \cite{gao2017} & 2.39 & \cite{fong2015} & 0.078 & 0.48 & -2.30 \\
            070714B & {\bf 0.01} & \cite{gao2017} & 2.30 & \cite{fong2015} & 0.078 & 0.48 & -0.30 \\
            080905A* & {\bf 0.007} & \cite{ascenzi2019} & 2.06 & \cite{fong2015} & 0.043 & 0.16 & -2.15 \\
            130603B & {\bf 0.03}, 0.01--0.1, 0.075 & \cite{jin2016} \cite{tanvir2013} \cite{ascenzi2019} & 2.70 & \cite{fong2015} & 0.056 & 0.94 & -1.00 \\
            150424A & {\bf 0.1} & \cite{ascenzi2019} & 2.30 & \cite{ascenzi2019} & 0.040 & 1.86 & -1.40 \\
            160821B* & {\bf 0.01}, 0.17, <0.006 & \cite{lamb2019}, \cite{ascenzi2019}, \cite{troja2019} & 2.30 & \cite{lamb2019} & 0.038 & 0.20 & -2.00 \\
            \hline
            150101B & >0.02, >0.1, 0.037 & \cite{troja2018} \cite{fong2016} \cite{ascenzi2019} & -- & -- & -- & -- & -- \\
            200522A & 0.1 & \cite{fong2021} & -- & -- & -- & -- & -- \\
        \end{tabularx}
* indicates a GRB where we reduced our jet model energy by a factor 10
\end{table}

\begin{figure}[H]
    \begin{adjustwidth}{-\extralength}{0cm}
    \centering
    \includegraphics[width=18cm]{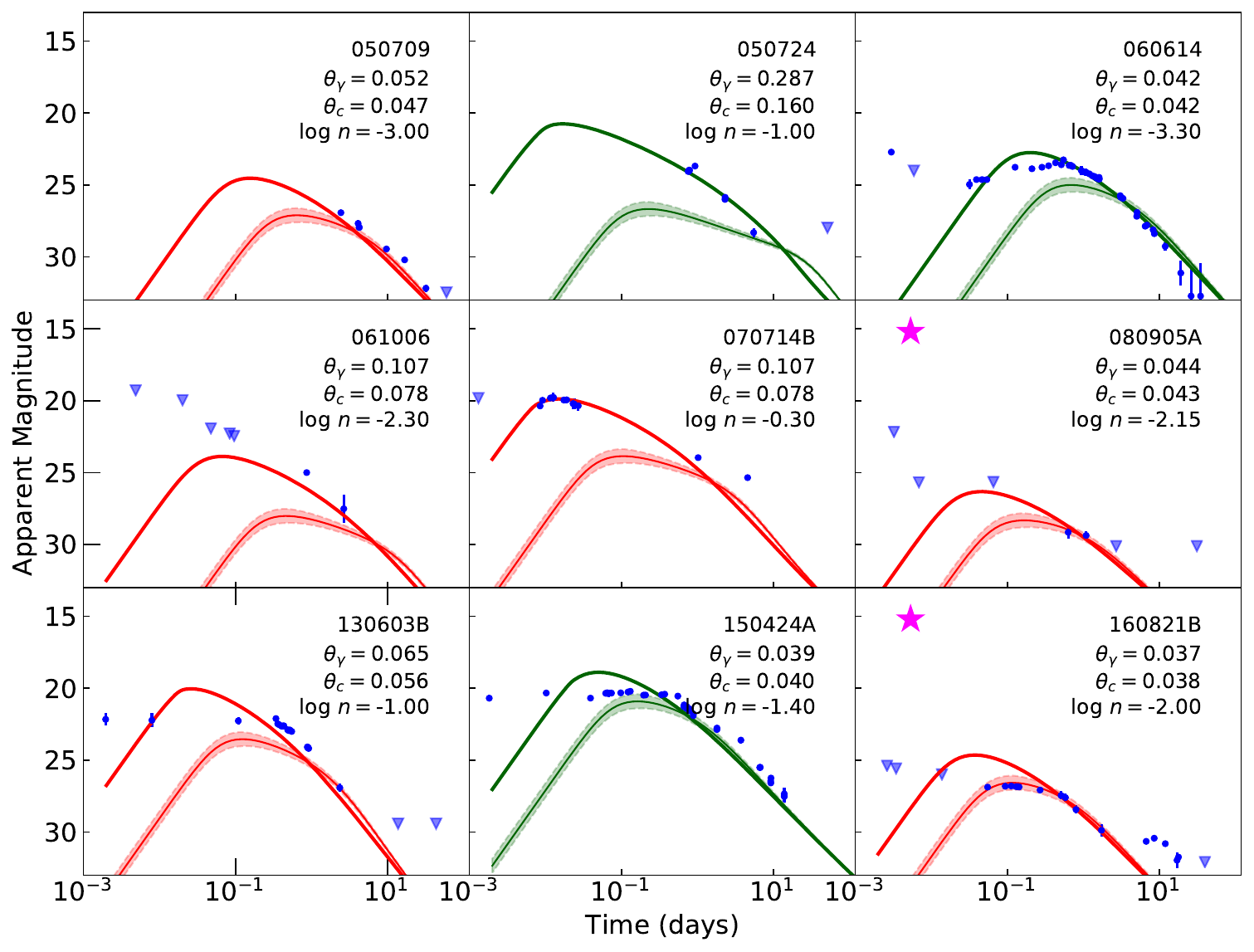}
    \end{adjustwidth}
    \caption{GRBs in our sample with candidate kilonova claims in the literature.
    Our afterglow model with a fixed energy is shown to be consistent with the observed lightcurves.
    In each case we modify the jet structure profile with the candidate kilonova's reported eject mass, and vary the ambient medium to give a satisfactory alignment between the $0\leq \iota \leq \theta_\gamma$ for the lightcurves, thick and thin red/green lines respectively. Lightcurves in green have $\varepsilon_B=0.1$ instead of the fiducial, $\varepsilon_B = 0.01$.
    Each panel shows $\theta_\gamma$, $\theta_c$, and the $\log~n$, where $n$ is given in cm$^{-3}$.
    Panels marked with a pink star (GRBs 080905A, and 160821B) have a reduced energy (by a factor 0.1) to better describe the observed afterglows.
    Note -- afterglow parameters are not fits to the data, and the afterglow models do not include the kilonova, reverse shock, or energy injection contribution.}
    \label{fig:LCs_AG}
\end{figure}

Figure \ref{fig:LCs_AG} shows the 9 kilonova candidate GRB afterglows in our sample.
For each GRB we use the ejecta mass value listed in bold from Table \ref{tab:KN-GRBs} to estimate { a new} jet core size using equation \ref{eq:theta_fit}. 
{ As we keep the jet core energy constant, changing the model core size will result in a different isotropic equivalent kinetic energy for the jet structure model profile.
The new isotropic equivalent kinetic energy, for the $\theta=0$ point, is found assuming the core energy in the jet from our `Averaged' model is conserved, so $E_{K,{\rm iso}}(\theta=0) = (1-\eta) 1.9\times10^{52} (1.79\times10^{-3}/\theta_c^2)$, where $\eta$ is the efficiency of the $\gamma$-ray emission, and $\theta_c$ is the new core angle.} 
{ These new values for $\theta_c$ and central isotropic equivalent energy are} then used to substitute the values in our `Averaged' jet structure profile shown in Figure \ref{fig:bootstrap}.

{ In Figure \ref{fig:LCs_AG}, model lightcurves shown with a red line} have
fixed microphysical parameters, $\sqrt\varepsilon_B = \varepsilon_e = 0.1$, and where the model lightcurve is shown with a green line, we change to $\varepsilon_B=0.1$ to give better agreement with the data.
{ For each lightcurve we use values for the index $p$ from literature, as listed in Table \ref{tab:KN-GRBs}}.
For each GRB afterglow we vary the value of the ambient density to find an approximate alignment between our fiducial model and the afterglow data\footnote{A sophisticated fit to these data sets would require the inclusion of reverse shock emission, refreshed shock or energy injection, plus a kilonova contribution. This is beyond the scope of this work, however, using the literature values of the index $p$, should ensure that our approximate lightcurve models are consistent with any X-ray afterglow flux for individual bursts. Other works have focused on fitting afterglow and kilonova models to data e.g. \cite{ascenzi2019}.}.
For GRBs 080905A, and 160821B, we could not find a satisfactory alignment using our total jet energy and varying only the ambient density.
However, as described in \cite{lamb2019}, GRB 160821B requires an initially low-power jet that is refreshed at later times\footnote{The data presented here for GRBs 130603B and 160821B has the kilonova dominated data removed \citep[see][]{tanvir2013, lamb2019}, however, all other GRBs in our sample may include multiple emission components and, potentially, a significant contribution from a kilonova.}.
The equivalent energy of the initial jet in GRB 160821B is $\sim0.1$ that of our fiducial jet, 
however, the energy injection at $\gtrsim1$ days for GRB 160821B results in a total energy that is consistent with the energy of our single episode jet model \citep[see][for details]{lamb2019}.
We apply this same reduction in initial jet energy to GRB 080905A to achieve a better alignment with the observed flux density.
Our model lightcurves do not include the refreshed shock contribution, however, for GRB 160821B this is equivalent to the late time `excess' seen in the data.
No data, other than the non-constraining upper-limits, at $\gtrsim10$ days is available for GRBs 080905A to test the refreshed shock scenario.

The model afterglows in Figure \ref{fig:LCs_AG} also show the maximum variability due to the rotational orientation of the jet with respect to the observer, as shown in Figure \ref{fig:DLFp} -- note that this uncertainty in peak flux does not affect the on-axis case.
The angle $\theta_\gamma$ provides an indication of the highest angle from which an observable GRB is likely to be emitted, however, this is not a hard limit and where beaming effects, the distance to the source, and secondary $\gamma$-ray emitting components are considered e.g. shock breakout of the cocoon, then the inclination at which an observer could detect the GRB is higher for nearby sources.

The lightcurves for GRBs { 050724,} 060614 and 150424A stand out in our sample as having a late break to the steep decline phase when compared with our fiducial models, and the green lightcurves shown in Figure \ref{fig:LCs_AG} have used $\varepsilon_B=0.1$ to give better agreement with the data -- additionally noting that the choice of initial jet energy in our simulation was arbitrary\footnote{The isotropic equivalent energy of the injected jets {(calculated assuming all the jet energy is uniformly within the core angle for the initially Gaussian structured jet) } are $\sim6\times10^{50}$ erg, and slightly lower than the $\sim10^{51}$ erg expected for mergers \citep{shapiro2017, fryer2019}.}.
The opening angle for these jet models relies on the ejecta mass estimates, and for GRB 050724, the value of $0.001$ M$_\odot$ is the smallest in our sample and may well be underestimated. 
GRB 060614 is technically a long-duration GRB, with a prompt burst episode lasting $\sim100$s, however, the absence of an accompanying bright supernova combined with it exhibiting an initial spike of gamma-rays with a duration of only a few seconds has led to speculation that it could have been produced by a compact binary merger \citep{galyam2006, gehrels2006, perley2009, kann2011}.
The differences in these GRBs \citep[noting that GRB~150424A which has a large uncertainty on the source redshift, e.g.,][]{knust2017} may indicate that these bursts have a different progenitor to typical short GRBs i.e. they may be the result of a neutron star-black hole merger as opposed to a binary neutron star merger \citep[e.g.][]{gompertz2020}, where the potential energy budget for the jet is marginally higher \citep[e.g.][]{shapiro2017} or the ejecta mass is lower, especially in the polar regions, resulting in a potentially wider emergent jet structure.

The compact stellar merger origin for short GRBs has a limited energy budget for the jets, typically of the order $\sim 10^{51}$ erg for neutron star mergers \citep{fryer2019}.
Yet the observed distribution for the energy in the short GRB population spans $\sim 4$ orders of magnitude, see Figure \ref{fig:ekinetic} and \cite{fong2015}.
Here we have shown that inhomogeneity in the energy and velocity distribution across an emergent jet's surface due to turbulent processes via jet-ejecta mixing can account for a significant fraction of the observed spread in inferred kinetic energies.
By taking into account the range of ejecta masses from the candidate kilonova population, the effective opening angle (the jet core angle in our notation) for a GRB jet with a fixed, or limited, energy budget can be determined.
When combined with the expected variation in the ambient density between sources we can explain much of the variety in afterglow flux and timescales for the small sample of GRBs with kilonova candidates.

Figure \ref{fig:prelim-lc}, where $\iota\sim 4-5 \theta_c$, and fits for the rotationally averaged simulation jet structure profiles in \cite{nativi2022} show that such a functional jet structure is additionally consistent with the afterglow of GW170817/GRB 170817A -- we do not repeat that analysis here. 
For the afterglow to GRB 170817A, multiple jet structure profiles have shown viable fits to the data, including more exotic structures than the core-dominated profiles presented here \citep[e.g.,][]{takahashi2021}, as well as the refreshed shock/energy injection scenario \citep{lamb2020}.
{Several authors have shown how the jet structure profiles used to explain the lightcurve of the afterglow of GRB 170817A resemble the cosmological population of short GRB afterglows when viewed at a very small inclination \citep[e.g.][]{salafia2019, fong2019, mogushi2019, tan2020}, thus renewing the case for a universal jet structure profile for merger jets \citep[note however that a refreshed shock jet model can also explain the observed temporal structure of the afterglow of GRB 170817A e.g.,][]{lamb2020}.
Here we expand on the general case for merger jets and give a physically motivated jet structure profile to use in such cases where an ad hoc or analytically derived profile has been used before.}
The parameters for this analytic profile include uncertainties due to the rotational variation within our simulated jet profiles, these can be used as a physically motivated prior on jet structure parameters when fitting observed data.
Additionally, with equation \ref{eq:theta_fit}, we include a way to adjust the jet core size given different assumptions on the injected jet's half-opening angle, $\theta_0$, and the jet energy, $E_j$.

\section{Conclusions}\label{sec:conc}

We have demonstrated that the inhomogeneity in energy and velocity across the jet surface of a 3D hydrodynamic jet-neutron star merger wind simulation results in a peak afterglow flux density that depends on the observer-system orientation relative to the jet central axis in terms of both polar and rotation angles.
The potential change in peak flux with orientation within a $\gamma$-ray emitting region of a jet with a fixed total energy results in:
\begin{itemize}
    \item Variation in peak afterglow flux density due to rotation, $<0.5$ dex.
    \item Variation in peak afterglow flux density due to inclination (polar orientation), $<1.3$ dex.
    \item An order of magnitude spread in jet kinetic energy distribution when inferred from the peak afterglow, where the $\gamma$-ray efficiency of the GRB emission is fixed.
\end{itemize}

We define a physically motivated analytic function for a typical neutron star merger jet, and demonstrate how, for a given jet energy and merger ejecta mass, the effective opening angle of the jet can change.
Using the literature reported ejecta masses for six candidate kilonovae, we show that a fixed injected jet energy ($E_{\rm iso}\sim 6\times10^{50}$ erg) with our analytic structure function can account for the diversity of observed afterglow lightcurves by changing only the ambient medium density.
For { two} GRB-kilonova candidates in our sample; GRBs 080905A, and 160821B, we find that the jet energy of our fiducial model must be reduced by an order of magnitude.
This reduction is consistent with the refreshed shock scenario used to describe the afterglow of GRB 160821B in \cite{lamb2019}, where the total, post-injection energy of the afterglow is equivalent to that of our fiducial model.
This indicates that the intrinsic energy range for the short GRB population is likely very narrow, and consistent with the theoretical prediction for neutron star merger central engines.

The analytic jet structure as given in equations \ref{eq:sj1}-\ref{eq:sj2} can be used with the parameters in Table \ref{tab:profiles} for a physically motivated jet structure profile to include a prior uncertainty on the functional parameters.
Additionally, this structure can be modified with equation \ref{eq:theta_fit}, to accommodate different merger ejected masses, jet energies, or injected jet opening angles, to replace the typically ad-hoc jet structure functions used in afterglow modelling.


\vspace{6pt} 



\authorcontributions{Conceptualization, G.P.L.; methodology, G.P.L., L.N., C.L. and S.R.; software, G.P.L., L.N., S.R. and C.L.; validation, C.L., L.N. and G.P.L.; formal analysis, L.N. and G.P.L.; resources, S.R., N.T. and A.L.; data curation, L.N., G.P.L. and D.A.K.; writing---original draft preparation, G.P.L; writing---review and editing, L.N., S.R., D.A.K., A.L. and N.T.; visualization, G.P.L. and L.N.}

\funding{GPL is supported by STFC grant  ST/S000453/1. DAK acknowledges support from Spanish National Research Project RTI2018-098104-J-I00 (GRBPhot). SR and LN were supported by the Swedish Research 
Council (VR) under grant number 2016\_03657, by the Swedish National Space Board under grant number
Dnr. 107/16, by the research environment grant  ``Gravitational Radiation and Electromagnetic 
Astrophysical Transients (GREAT)'' funded by the  Swedish Research council (VR) under 
Dnr 2016\_06012 and by the Knut and Alice Wallenberg Foundation  under grant Dnr. KAW 2019.0112.
SR has further been supported by the Deutsche Forschungsgemeinschaft (DFG, German Research Foundation) 
under Germany’s Excellence Strategy -EXC 2121- "Quantum Universe" - 390833306 and by the European 
Research Council (ERC) Advanced Grant INSPIRATION under the European Union's Horizon 2020 
research and innovation programme (Grant agreement No. 101053985). The hydrodynamic simulations 
were performed on resources provided by the Swedish National Infrastructure for
Computing (SNIC) at Beskow and on the facilities of the North-German Supercomputing Alliance (HLRN).}

\dataavailability{Data available on reasonable request.
} 

\acknowledgments{GPL thanks Fergus Hayes, En-Tzu Lin, Laurence Datrier, Surojit Saha, Michael Williams, Siong Heng, Martin Hendry, Albert Kong, and John Veitch for useful discussions via online meetings throughout 2020-2021.}

\conflictsofinterest{The authors declare no conflict of interest. The funders had no role in the design of the study; in the collection, analyses, or interpretation of data; in the writing of the manuscript; or in the decision to publish the results.} 



\abbreviations{Abbreviations}{
The following abbreviations are used in this manuscript:\\

\noindent 
\begin{tabular}{@{}ll}
GRB & Gamma-ray Burst
\end{tabular}
}


\begin{adjustwidth}{-\extralength}{0cm}

\reftitle{References}


\bibliography{ms.bib}

%


\end{adjustwidth}
\end{document}